\documentstyle[sprocl,epsfig]{article}

\def\be{\begin{equation}}
\def\ee{\end{equation}}
\def\bea{\begin{eqnarray}}
\def\eea{\end{eqnarray}}

\newcommand{\AmS}{{\protect\the\textfont2
  A\kern-.1667em\lower.5ex\hbox{M}\kern-.125emS}}
\newcommand{\lsim}{\raisebox{-3pt}{$\,\stackrel{\textstyle <}{\sim}\,$}}
\newcommand{\gsim}{\raisebox{-3pt}{$\,\stackrel{\textstyle >}{\sim}\,$}}

\def\Pom{{\bf I\!P}}

\def\lsim{\mathrel{\rlap{\lower4pt\hbox{\hskip1pt$\sim$}}
    \raise1pt\hbox{$<$}}}         

\def\gsim{\mathrel{\rlap{\lower4pt\hbox{\hskip1pt$\sim$}}
    \raise1pt\hbox{$>$}}}         

\begin{document}

\title{
HELICITY FLIP IN DIFFRACTIVE DIS }



\author{N.N. Nikolaev}

\address{Institut f. Kernphysik, Forschungszentrum J\"ulich,
D-52450 J\"ulich, Germany\\
 L.D.Landau Institute for Theoretical Physics,
142432 Chernogolovka, Russia
\\E-mail: N.Nikolaev@fz-juelich.de } 


\maketitle\abstracts{ 
I review new results on $s$-channel 
helicity nonconservation (SCHNC) in diffractive DIS. I
discuss how by virtue of unitarity diffractive DIS gives rise to 
spin structure functions which were believed to vanish at $x\ll 1$.
These include tensor polarization of sea quarks in the deuteron,
strong breaking of the Wandzura-Wilczek relation and demise of 
 the Burkhardt-Cottingham sum rule.\\
The invited talk at QCD \& Multiparticle Production, XXIX International 
Symposium on Multiparticle Dynamics (ISMD99), August 9-13, 1999. 
Brown University, Providence, RI  02912, USA }

\section{Introduction}

The often repeated argument is that diffractive scattering is driven 
via unitarity by absorption due to multiparticle production. 
The common wisdom is that this entails vanishing spin-dependence 
of diffractive scattering. The related 
QCD argument has been that quark helicity 
conservation entail the $s$-channel 
helicity conservation (SCHC) at small $x$, i.e., decoupling of  
QCD pomeron from helicity flip. Here I review the recent discovery 
\cite{NZDIS97,NPZLT,KNZ98,IN99,AINP,G2} of substantial SCHNC  
in diffractive DIS into the both continuum and vector mesons. 
Furthermore, SCHNC diffractive DIS in conjunction with unitarity 
changes dramatically the small-$x$ behaviour of spin structure function
$g_{2}$ and leads to the demise of the Burkhardt-Cottingham sum rule and 
the departure from the Wandzura-Wilczek relation. Still another
diffraction driven spin effect is the tensor polarization of sea
quarks in the deuteron.

\section{Is SCHNC compatible with quark helicity conservation?}

The backbone of DIS  is the Compton scattering (CS) $\gamma^{*}_{\mu}p\to 
\gamma^{*'}_{\nu}p'$. 
The CS amplitude $A_{\nu\mu}$ can be written as 
$
A_{\nu\mu}=\Psi^{*}_{\nu,\lambda\bar{\lambda}}\otimes A_{q\bar{q}}\otimes
\Psi_{\mu,\lambda\bar{\lambda}}
$
where $\lambda,\bar{\lambda}$ stands for $q,\bar{q}$ helicities,
$\Psi_{\mu,\lambda\bar{\lambda}}$ is the wave function of the $q\bar{q}$ 
Fock state of the photon. The QCD pomeron exchange
$q\bar{q}$-proton scattering
kernel $A_{q\bar{q}}$  does not depend on, and conserves exactly, 
the $q,\bar{q}$ helicities. For
nonrelativistic massive quarks, $m_{f}^{2} \gg Q^{2}$, one only has
transitions $\gamma^{*}_{\mu} \to q_{\lambda} +\bar{q}_{\bar{\lambda}}$ 
with $\lambda +\bar{\lambda}=\mu$. However, 
the relativistic P-waves give rise to 
transitions
of transverse photons $\gamma^{*}_{\pm}$ into the $q\bar{q}$ state with 
$\lambda +\bar{\lambda}=0$
in which the helicity of the photon is transferred to the $q\bar{q}$ orbital 
 momentum. Consequently, the QCD pomeron exchange SCHNC transitions
$\gamma^{*}_{\pm} \to (q\bar{q})_{\lambda +\bar{\lambda}=0} \to
\gamma^{*}_{L} ~~~{\rm and}~~~
\gamma^{*}_{\pm} \to (q\bar{q})_{\lambda +\bar{\lambda}=0 }\to
\gamma^{*}_{\mp} $ 
are allowed \cite{NZDIS97,NPZLT} and SCHNC persists at small $x$. 
We emphasize that the above argument 
for SCHNC does not require applicability of pQCD.

\begin{figure}
\epsfysize 4 in 
\epsfbox{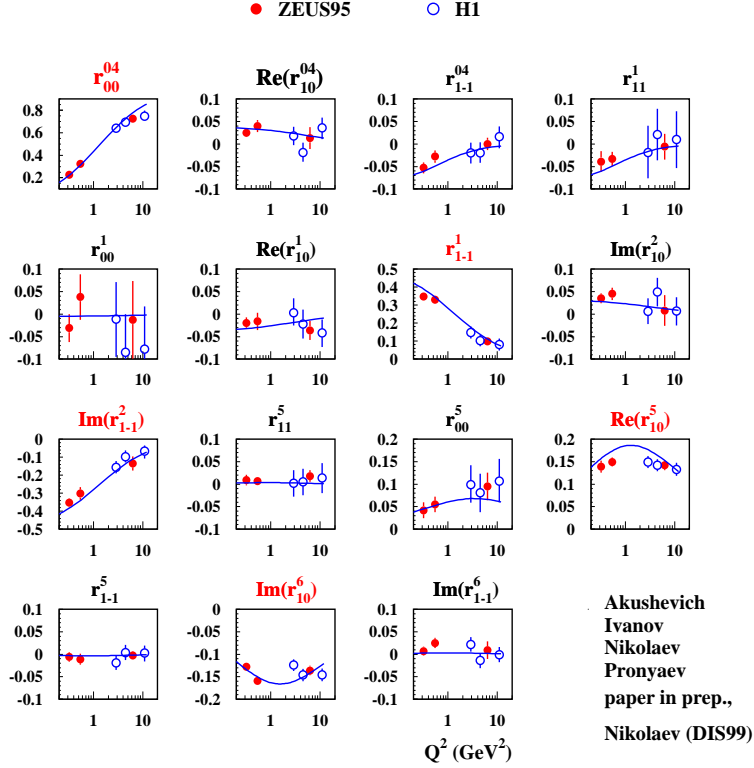}
\caption{Our prediction \protect\cite{AINP} 
for the 
spin density matrix $r_{ik}^{n}$ 
of diffractive $\rho^{0}$-meson vs. the experimental
data from
ZEUS \protect\cite{ZEUS} and H1 \protect\cite{H1}.}
\label{fig:1}
\end{figure}

\section{LT interference and SCHNC in diffractive DIS }

The first ever direct evaluation \cite{NZDIS97} of SCHNC effect in QCD - the
 LT-interference 
of transitions $\gamma^{*}_{L}p \to p'X$ and $
\gamma^{*}_{\pm}p \to p'X$ into the same continuum 
diffractive states $X$  - has been reported in 1997.
 Experimentally, it can be measured at HERA by both H1 and
ZEUS via azimuthal correlation between the $(e,e')$ and 
$(p,p')$ scattering planes and can be used the determination of the otherwise elusive 
$R=\sigma_{L}/\sigma_{T}$ for diffractive DIS is found in \cite{NPZLT}. 
The principal issue is that this asymmetry persists, and even rises slowly, at small 
$x_{\Pom}$.

 The azimuthal correlation
of the $(e,e')$ and $(p,V)$ planes with the vector meson decay 
plane, and the in-decay-plane
angular distributions of decay products, allow the experimental
determination of all helicity amplitudes $A_{\mu\nu}$ for
diffractive $\gamma^{*}p\to Vp'$.
One finds that helicity flip only is possible due to 
the transverse and/or longitudinal Fermi motion of quarks and
is extremely sensitive to spin-orbit coupling in the
vector meson, I refer for details to \cite{KNZ98,IN99}.
The consistent analysis of production of the 
$S$-wave and $D$-wave vector mesons is presented only
in \cite{IN99}.  One would readily argue that by exclusive-inclusive 
duality \cite{GNZlong} between diffractive
 DIS into continuum and vector mesons \cite{NZDIS97,NPZLT} the dominant 
SCHNC effect in vector meson production is the interference 
of SCHC $\gamma^{*}_{L} \to V_L$ and SCHNC $\gamma^{*}_{T}\to V_L$ production,
i.e., the element $r_{00}^{5}$ of the vector meson spin
density matrix. The overall agreement
between our theoretical estimates 
\cite{AINP} of the spin density matrix $r_{ik}^{n}$ for
diffractive  $\rho^{0}$ 
and the ZEUS \cite{ZEUS} and H1 \cite{H1} experimental data
is very good. There is a clear evidence for  $r_{00}^{5}\neq 0$.

 In the $D$-wave state the total spin 
of $q\bar{q}$ pair is predominantly opposite to the spin of the $D$-wave 
vector meson. As a results, SCHNC in production of $D$-wave vector
mesons is much stronger \cite{IN99} than for the ground state S-wave mesons, which 
may facilitate the disputed $D$-wave 
vs. $2S$-wave assignment of the $\rho'(1480)$ 
and $\rho'(1700)$ and of the $\omega'(1420)$ and $\omega'(1600)$.
Striking predictions for D-wave meson production include  abnormally large 
higher twist corrections \cite{IN99} and non-monotonous  $Q^2$ 
dependence of  
$R^{D}=\sigma_{L}/\sigma_{T}$.

\section{Impact of diffraction upon $g_{2}$: breaking of the
Wandzura-Wilczek relation and demise of the Burkhardt-Cottingham sum rule }

The transverse spin asymmetry $A_{2}$ in polarized DIS is proportional to the
amplitude of forward CS 
$\gamma^{*}_{L}p\!\!\uparrow \to \gamma^{*}_{T}p\!\!\downarrow$ which is 
proportional to $g_{LT}=g_{1}+g_{2}$. In the standard two-gluon $t$-channel 
tower approximation the cross-talk
of the target and beam helicity flip only is possible at
the expense of suppression of
the small-$x$ behaviour of $x^2g_{LT}$  by the
extra factor $\sim x$ compared to $F_{1}$, because  
not both gluons in the pomeron can have the nonsense polarization simultaneously.

The more familiar argument for the vanishing 
$A_{2}$ has been the parton model Wandzura-Wilczek relation between 
$g_{LT}$ and $g_{1}$ \cite{WW}:
$$
g_{LT}(x,Q^{2})=\int_{x}^{1} {dy \over y} g_{1}(x,Q^{2})\, ,
$$
which entails $g_{LT} \sim g_{1}$.    
Because the diffractive pomeron exchange does not contribute to $g_{1}$, 
the WW relation can be reinterpreted as a vanishing pomeron exchange
contribution to $g_{LT}$. This vanishing of the pomeron contribution 
has been the principal motivation behind
the much discussed Burkhardt-Cottingham \cite{BC} (BC) sum rule
$\int_{0}^{1} dx g_{2}(x,Q^{2})=0$. 

\begin{figure}
\epsfysize 1.7 in 
\epsfbox{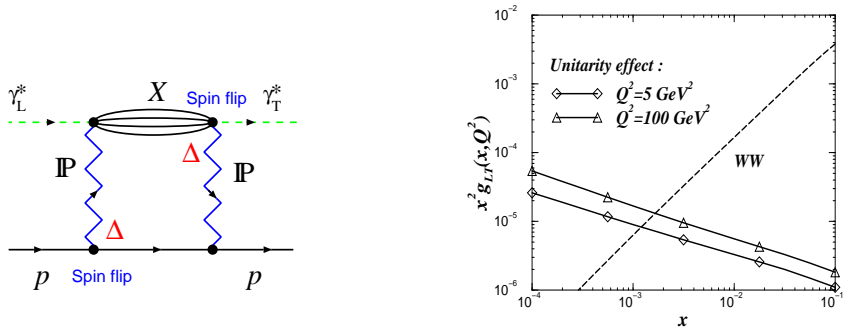}
\caption{The LHS: The unitarity diagram for diffractive contribution to
$g_{LT}$. The RHS: The unitarity correction to, and WW relation based 
evaluation of, $g_{LT}$. }
\label{fig:2}
\end{figure}

Our recent discovery is that diffractive SCHNC destroys via unitarity 
the both WW relation and BC sum rule \cite{G2}. The unitarity diffractive 
contribution (fig.~2) to CS has as building blocks 
diffractive amplitudes $\gamma^{*}p \to p'X$ in which there is
a helicity flip sequence, $\gamma^{*}_L \to X_{L} \to \gamma^{*}_T$ in the top 
blob and helicity flip sequences either 
$p\!\!\uparrow \to p'\!\!\uparrow \to p\!\!\downarrow$ 
or $p\!\!\uparrow \to p'\!\!\downarrow \to p\!\!\downarrow$
in the bottom blob. The both blobes are proportional to ${\bf \Delta}$
and vanish for forward produced $X$, but upon the integration 
over the phase space of $p'X$ one finds the nonvanishing 
$\int d^{2}{\bf \Delta} \Delta_{i}\Delta_{k}$ and unitarity driven transition 
$\gamma^{*}_{L}p\!\!\uparrow \to \gamma^{*}_{T}p\!\!\downarrow$ which does
not vanish in the forward direction. The principal point 
is that the unitarity diagram furnishes the
cross-talk of the beam and target helicity flip with pure nonsense
polarizations of all the four exchanged gluons in the two pomerons.

Our result \cite{G2} for the diffraction-driven $g_{LT}$ reads 
\begin{equation}
g_{LT}(x,Q^{2}) \propto {1\over x^{2}} r_{5}\int_{x}^{1} {d\beta\over \beta}
g_{LT}^{D}(x_{\Pom}={x\over \beta},Q^{2})\, .
\end{equation}
It rises steeply at small $x$. It is the scaling function of $Q^{2}$ because 
the diffractive LT structure function $g_{LT}^D(x_{\Pom},Q^{2})$ is the
scaling one. The resulting 
 asymmetry $A_{2} \propto xg_{LT}/F_{1}$ does not vanish 
at small $x$, furthermore, at a moderately small $x$ it even rises because
$g_{LT}^D(x_{\Pom},Q^{2}) \propto G^{2}(x_{\Pom},\overline{Q}^{2})$ where
$G(x_{\Pom},\overline{Q}^{2})$ is the gluon SF of the proton and
$\overline{Q}^{2} \sim$ 0.5-1 GeV$^2$. 

In fig.~3 we show how the steeply rising unitarity 
correction overtakes at small $x$ the standard  $g_{LT}$ evaluated from the 
Wandzura-Wilczek 
(WW) relation starting with fits to the world data on $g_{1}$. 
As such our unitarity
effect is the first nontrivial scaling departure from the WW relation

Finally, the above breaking of the WW relation implies $g_{LT} \gg g_{1}$ 
and $g_{2} = g_{LT}$ at very small $x$. Consequently, the unitarity-driven
rise of $g_{2}$ 
destroys the BC sum rule because the BC integral would diverge severely.
Incidentally, the BC sum rule has always been suspect.

\section{Diffraction, unitarity and tensor structure function of the deuteron} 

\begin{figure}
\epsfysize 1.8 in 
\epsfbox{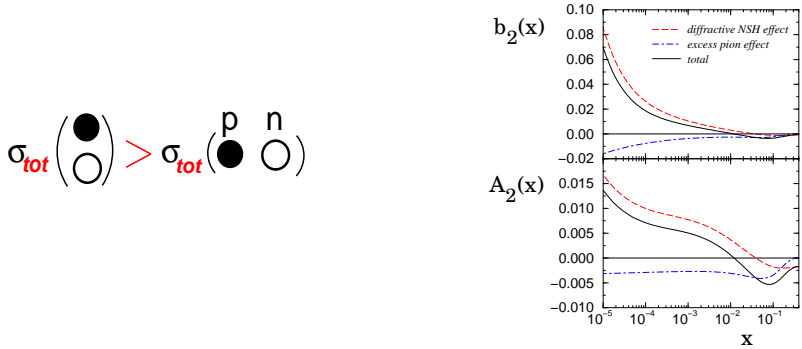}
\caption{The LHS: The unitarity diagram for diffractive contribution to
$g_{LT}$. The RHS: The unitarity correction to, and WW relation based 
evaluation of, $g_{LT}$. }
\label{fig:3}
\end{figure}

The deuteron is an unique spin-1 target. Because of the S-D wave interference
it is a dumbbell and spatial orientation of nucleons and their Fermi
motion in the deuteron depend 
on its polarization which may give rise to the dependence of parton
densities on the tensor polarization of the deuteron. In the impulse
approximation the tensor structure function $b_{2}(x)$ is found to be
negligible, per mill,  effect \cite{Jaffe} and satisfies the 
Close-Kumano sum rule \cite{Close} $\int_{0}^{1} {dx \over x} b_{2}(x)=0.$
Close and Kumano conjectured this is a generic QCD sum rule.

Fig.~3 makes it obvious that diffractive eclipse effect in the deuteron 
depends on its orientation which is controlled by tensor polarization. 
The results of the calculation \cite{TensorD} 
of the eclipse effect are shown in fig.~3. The tensor polarization of
the sea measured by tensor asymmetry $A_{2}$ is quite substantial and
persists at small $x$. Apart from diffractive eclipse effect, it receives 
a small contribution also from the pion exchange \cite{TensorD}.

\section*{Conclusions}

Drastic revision of our prejudice on spin-independence of diffraction
is called upon.
The recent fundamental finding is that 
QCD pomeron exchange does not conserve the $s$-channel helicity.
The mechanism of SCHNC is well understood. SCHNC offers an unique 
window at the spin-orbit coupling in vector mesons. SCHNC in diffractive
DIS drives, via unitarity relation, a dramatic small-$x$ rise of the transverse
spin structure function $g_2$ which breaks the Wandzura-Wilczek relation
and invalidates the Burkhardt-Cottingam sum rule. The related unitarity
effect is the tensor polarization of sea quarks in the deuteron which persists at
small $x$. \\

\noindent
{\bf Acknowledgments.} I'm indebted to C.-I.Tan  for the
invitation to ISMD-99.

\end{document}